\documentclass[a4paper,fleqn,usenatbib]{mnras}

\usepackage{newtxtext,newtxmath}

\usepackage[T1]{fontenc}
\usepackage{ae,aecompl}
\usepackage{ulem} 

\usepackage{graphicx}	
\usepackage{amsmath}	

\title[GeV spectral break of 3C 454.3]{On the origin of GeV spectral break for Fermi blazars: 3C 454.3}

\author[S. J. Kang et al.]{
Shi-Ju Kang,$^{1,2}$\thanks{E-mail: kangshiju@alumni.hust.edu.cn}
Yong-Gang Zheng,$^{2}$\thanks{E-mail: ynzyg@ynu.edu.cn}
Qingwen Wu,$^{3}$\thanks{E-mail: qwwu@hust.edu.cn}
Liang Chen,$^{4}$
and Yue Yin$^{1}$
\\
$^{1}$School of Physics and Electrical Engineering, Liupanshui Normal University, Liupanshui, Guizhou, 553004, China\\
$^{2}$Department of Physics, Yunnan Normal University, Kunming, Yunnan, 650092, China\\
$^{3}$School of Physics, Huazhong University of Science and Technology, Wuhan, Hubei, 430074, China\\
$^{4}$ Key Laboratory for Research in Galaxies and Cosmology, Shanghai Astronomical Observatory, Chinese Academy of Sciences, \\
~~~80 Nandan Road, Shanghai 200030, China
}

\date{Accepted 2021 February 16. Received 2021 January 13; in original form 2020 June 21}

\pubyear{2021}

\begin{document}
\label{firstpage}
\pagerange{\pageref{firstpage}--\pageref{lastpage}}
\maketitle

\begin{abstract}

The GeV break in spectra of the blazar 3C 454.3 is a special observation feature that has been discovered by the {\it Fermi}-LAT. The origin of the GeV break in the spectra is still under debate. In order to explore the possible source of GeV spectral break in 3C 454.3, a one-zone homogeneous leptonic jet model, as well as the {\it McFit} technique are utilized for fitting the quasi-simultaneous multi-waveband spectral energy distribution (SED) of 3C 454.3. The outside border of the broad-line region (BLR) and inner dust torus are chosen to contribute radiation in the model as external, seed photons to the external-Compton process, considering the observed $\gamma$-ray radiation. The combination of two components, namely the Compton-scattered BLR and dust torus radiation, assuming a broken power-law distribution of emitted particles, provides a proper fitting to the multi-waveband SED of 3C 454.3 detected 2008 Aug 3 - Sept 2 and explains the GeV spectral break. We propose that the spectral break of 3C 454.3 may originate from an inherent break in the energy distribution of the emitted particles and the Klein-Nishina effect. A comparison is performed between the energy density of  the ‘external’ photon field for the whole BLR $U_{\rm BLR}$ achieved via model fitting and that constrained from the BLR data. The distance from the position of the $\gamma$-ray radiation area of 3C 454.3 to the central black hole could be constrained at $\sim 0.78$pc ($\sim 4.00 R_{\rm BLR}$, the size of the BLR).
\end{abstract}



\section{Introduction}\label{section1}

Blazars are a class of active galactic nuclei (AGN) with a relativistic jet pointed within a small observation angle to the line of sight 
that can be subdivided into flat-spectrum radio quasars (FSRQs) and BL Lacertae objects (BL Lacs) \citep{1995PASP..107..803U}.
The multi-wavelength spectral energy distribution (SED) from the radio to the $\gamma$-ray bands of blazars 
is mostly caused by the non-thermal radiation, where the SED generally exhibits a two-bump framework in
the ${\rm log}{\nu}-{\rm log}{\nu}{F}_{\nu}$ space. 
Based on the peak frequency ($\nu^{\rm S}_{\rm p}$) of the first hump, 
blazars are categorized as low (LSP, e.g., $\nu^{\rm S}_{\rm p}<10^{14}$Hz), 
intermediate (ISP, e.g., $10^{14}\rm Hz<\nu^{\rm S}_{\rm p}<10^{15}$Hz) 
and high-synchrotron-peaked (HSP, e.g., $\nu^{\rm S}_{\rm p}>10^{15}$Hz) blazars (e.g, \citealt{2010ApJ...716...30A}).

It can usually be confirmed that the lower energy bump is generally related to 
the synchrotron (Syn) radiation generated by the jet's non-thermal electrons  \citep{1998AdSpR..21...89U}. 
However, the origin of the second bump could be considered an open problem. 
In the leptonic model schemes, the second bump ($\gamma$-ray) generally is induced by inverse Compton scattering. 
The seed photons for inverse Compton scattering may be derived from synchrotron photons in the jet (synchrotron self-Compton, SSC, process, e.g., \citealt{1981ApJ...243..700K,1985ApJ...298..114M,1989ApJ...340..181G,1992ApJ...397L...5M,2003ApJ...597..851K}),
and/or the external photons (external-Compton, EC, process) from outside the jet
(e.g., \citealt{1993ApJ...416..458D,1994ApJ...421..153S,1996MNRAS.280...67G,1998ApJ...501L..51B}).
Consider that these external photons can be due to the accretion disk (e.g., \citealt{1993ApJ...416..458D}),
the broad-line region (BLR) (e.g., \citealt{1994ApJ...421..153S,1996MNRAS.280...67G}), 
and the molecular torus (e.g., \citealt{2000ApJ...545..107B,2008MNRAS.387.1669G}). 
The SSC model has been extensively employed for fitting the multi-wavelength SED of HSP BL Lacs 
(e.g., \citealt{2004ApJ...601..151K,2012ApJ...752..157Z}),
while FSRQs are often better described by an SSC+EC
(e.g., \citealt{2002ApJ...581..127B,2011MNRAS.414.2674G,2014MNRAS.439.2933Y,2014ApJ...788..104Z,2019NewAR..8701541H}).

The radio source 3C 454.3 is a famous FSRQ at redshift z = 0.859 in the {\it Fermi Gamma-Ray Space Telescope} Large Area Telescope ({\it Fermi}-LAT) point source catalogs, named as 1FGL J2253.9+1608, 2FGL J2253.9+1609, 3FGL J2254.0+1608, and 4FGL J2253.9+1609 in the First, Second, Third, and Fourth {\it Fermi}-LAT catalogs (1FGL; \citealt{2010ApJS..188..405A}, 2FGL; \citealt{2012ApJS..199...31N}, 3FGL; \citealt{2015ApJS..218...23A}, and 4FGL; \citealt{2020ApJS..247...33A}), respectively. 
Since the {\it Fermi Gamma Ray Space Telescope} was launched on June 11, 2008, a new era for studying $\gamma$-ray astronomy has been started. 
An interesting feature of the $\gamma$-ray spectrum for 3C 454.3 is the GeV spectral break, which was first detected by {\it Fermi}-LAT in \cite{2009ApJ...699..817A,2010ApJ...710.1271A}, 
and was further confirmed subsequently by \cite{2011ApJ...743..171A,2015ApJ...810...14A} and \cite{2014ApJ...794....8S}. 
The $\gamma$-ray spectrum is characterized by a break around 2 GeV with photon indices of 
$\Gamma_1 \simeq 2.3 \pm 0.1$  and $\Gamma_2 \simeq 3.5 \pm 0.25$
that are lower and higher than the break energy, respectively \citep{2009ApJ...699..817A}.

This break is known to occur in both bright FSRQs, and in several LSP BL Lacs (e.g., \citealt{2010ApJ...710.1271A}).  
The GeV break energies tend to occur in the 1-10 GeV range (e.g., \citealt{2010ApJ...710.1271A,2010ApJ...717L.118P,2012ApJ...761....2H}), 
and are relatively stable (e.g., \citealt{2010ApJ...721.1383A,2011ApJ...733L..26A,2011MNRAS.417L..11S}).
Such as, the break energy is basically constant ($\sim$ a few GeV) in the flaring state for some sources, specifically for 3C 454.3 \citep{2011ApJ...733L..26A,2014ApJ...790...45P}.
A minor change in the spectral break energy for significant deviations in flux conditions has been detected for 3C 454.3 \citep{2010ApJ...721.1383A}. 
According to the studies performed by \cite{2012ApJ...761....2H},
the average spectrum of 3C 454.3 could be optimally fit via a broken power-law function, 
regardless of the change in the break energy value in accordance with the energy range that fitting is accomplished.

Nowadays, the origin of the GeV break in Fermi blazar spectra (e.g., 3C 454.3) is still a puzzling and unresolved problem.
Several comments that have been presented to address this property are $\gamma-\gamma$ annihilation from He II line photons \citep{2010ApJ...717L.118P}, 
inherent electron spectral breaks \citep{2009ApJ...699..817A}, 
Ly-$\alpha$ dissipation \citep{2010ApJ...721.1383A},
Klein-Nishina (KN) influences occurred when BLR emission is scattered by jet electrons in a near-equipartition method  \citep{2013ApJ...771L...4C}, 
or a hybrid scattering (e.g., \citealt{2010ApJ...714L.303F,2013ApJ...771L...4C,2016A&A...589A..96H}).
For instance, the origin of the spectral break in 3C 454.3 has been discussed by \cite{2009ApJ...699..817A}. 
They proposed that the break in the $\gamma$-ray spectrum is most possibly caused by an intrinsic break in the energy distribution of the emitting particles near the electron energies $E_{\rm el}\sim10^3 m_e c^2$ (\citealt{2009ApJ...699..817A}).

The spectral break can not be due to photon interactions with a local emission field \citep{2009ApJ...699..817A}
that needs a  jet bulk Lorentz coefficient lower than that derived from superluminal radio viewings of 3C 454.3 ($\Gamma_{\rm bulk} \simeq 15$, \citealt{2005AJ....130.1418J}), 
or the extragalactic background light (EBL) due to the transparency of the universe to 40 GeV photons at $z \lesssim 1.0$  for all EBL models 
(e.g., \citealt{2006ApJ...648..774S,2009MNRAS.399.1694G,2010ApJ...712..238F}). 
Accordingly, the most reasonable statement is that the spectral break origin is an inherent break in the electron distribution \citep{2009ApJ...699..817A}. 
However, this statement may fail if the SED's optical/UV slope  is described with non-thermal synchrotron emission. 
\cite{2010ApJ...714L.303F} proposed that combining  two Compton scattering parts 
(with external photons from the accretion disk and BLR) can describe the break origin in the {\it Fermi}-LAT spectrum of 3C 454.3. 
The mentioned problem was also further discussed by \cite{2016A&A...589A..96H} using
a continuous, time-dependent injection, jet model based on Compton scattering of the external target emission fields
of the accretion disk and/or BLR, where the radiation area has been considered inside or beyond the BLR in the model fitting (e.g., \citealt{2016A&A...589A..96H}).
Also, according to \cite{2013ApJ...771L...4C} findings, the GeV spectral break could be generated by the Klein-Nishina effect when BLR photons are Compton scattered by a non-thermal, log-parabolic distributed population of electrons in the jet.

Since the position of the $\gamma$-ray radiation area in blazars is currently obscure, and could be considered an unresolved issue.
The radiation area has been generally considered inside or beyond the BLR in the model fitting, where the seed photons may originate from the accretion disk and/or the BLR and/or the dust torus (e.g., \citealt{2014ApJ...782...82D,2016ApJ...830...94F,2016A&A...589A..96H} and reference in).
We assume that the EC caused by the disk emission is negligible in 3C 454.3 since it is valid for at least some other bright quasars (e.g., in 3C 279, see \citealt{2018ApJ...853....6L}).
In the current study, we attempt to determine if the conventional one zone, leptonic model, containing numerous EC parts, can illustrate the GeV spectral break in 3C 454.3.
In this model, the external, soft photons originate in the BLR and dust torus and then are up-scattered through inverse Compton interactions with the non-thermal population of electrons, represented by a broken power-law distribution function.

The following cosmology is considered throughout
 $H_{0}=70\ \rm km\ s^{-1} Mpc^{-1}$, $\Omega_{0}=0.3$ and $\Omega_{\Lambda}=0.7$.

 \section{the model}\label{section2}

In the current study, the conventional one-zone synchrotron + inverse Compton model is employed to fit the SED of 3C 454.3, 
a model that has been extensively utilized in blazars (see \citealt{2010MNRAS.402..497G} and references in). 
The model has been employed in our previous works (see \citealt{2014ApJS..215....5K,2016MNRAS.461.1862K,2017ApJ...837...38K}). 
A homogeneous sphere with radius R installed in a magnetic field B is considered.
The homogenous sphere moves relativistically outward with the flow of the jet
with a speed  of $\upsilon=\beta c$ (c is the light speed in a vacuum, bulk Lorentz coefficient $\Gamma=1/\sqrt{1-\beta^2}$)
along the jet direction. 
The Doppler factor $\delta=\left[\Gamma\left(1-\beta\cos\theta\right)\right]^{-1}\approx\Gamma$ is selected for the relativistic jet with a narrow 
observation angle $\theta\leq 1/\Gamma$. The electron spectrum is chosen as a broken power-law distribution, with indices $p_{1}$ and $p_{2}$,
lower and higher than the break energy $\gamma_{b}m_{e}c^{2}$, respectively
\begin{equation}
   N(\gamma )=\left\{\begin{array}{ll}
            N_{0}\gamma ^{-p_1},  &  \mbox{ $\gamma_{\rm min}\leq \gamma \leq \gamma_{\rm b}$} \\
            N_{0}\gamma _{\rm b}^{p_2-p_1} \gamma ^{-p_2}, &  \mbox{ $\gamma _{\rm b}<\gamma\leq\gamma_{\rm max}$}
           \end{array},
           \right.
  \label{Ngamma}
  \end{equation}
where $\gamma_{\rm min}$ is the minimum Lorentz factor of electrons, $\gamma_{\rm max}$ is the maximum Lorentz factor of electrons, $\gamma_{\rm b}$ is the break Lorentz factor for the electron distribution, which follows from the balance between escape and cooling, and $N_{0}$ is the density normalization coefficient.  It is interesting to note that the broken power-law distribution might might result from the electrons with a power-law distribution being injected into the downstream flow they experience to escape and cool (see \citealt{2018MNRAS.478.3855Z} and references in).

Besides, an electron spectrum with a log-parabola distribution (see \citealt{2004A&A...413..489M,2004A&A...422..103M,2006A&A...448..861M}) produced by stochastic acceleration (e.g., \citealt{2009A&A...501..879T,2011ApJ...739...66T,2014ApJ...788..179C}) is given by
\begin{equation}
N(\gamma)=K_{0}\left({\frac{\gamma}{\gamma_{0}}}\right)^{-s-{b{\,}{\rm log}}\left(\frac{\gamma}{\gamma_{0}}\right)},
\end{equation}
where $s$ is the spectral index, $b$ is the spectral curvature parameter, $\gamma_{0}$ is the  peak Lorentz factor,  and $K_{0}$ is the normalization constant of a log parabola shape electron spectrum.

According to newly published studies, the $\gamma$-ray radiation area of blazar jets should be placed at about the outer border of (or beyond) the BLR and in the dust tours 
(e.g., \citealt{2016ApJ...821..102B,2017ApJ...837...38K,2017MNRAS.470.3283S,2017ApJS..228....1Z,2018A&A...619A.159M,2019A&A...627A.159H,2020A&A...634A.112T}),
where portions of both BLR and torus photons should be provided to illustrate the detected $\gamma$-ray spectrum. 
In the EC procedure, the external, soft photons are derived from two radiation zones 
(e.g., both BLR and dusty torus, \citealt{2013ApJ...771L...4C,2014ApJ...782...82D,2015MNRAS.454.1310Y,2015ApJ...803...15P,2017ApJ...837...38K}). 
Due to lack of clarity of $\gamma$-ray radiation area place, 
different from the assumptions of \cite{2010ApJ...714L.303F} and \cite{2016A&A...589A..96H}, 
they considered that the external, seed photons are originated from the accretion disk and BLR. 
The external, seed photons originate from both the dust torus and the whole BLR considered in this work.
As the same procedure performed by \cite{2013ApJ...771L...4C}, a dual-part Compton-scattering approach is chosen in which the external, seed photons generally originate from both the BLR and the dust torus. 
It is assumed that the $\gamma$-ray radiation area is located out of the BLR and within the dusty torus 
(e.g., \citealt{2016ApJ...821..102B,2017ApJ...837...38K,2017MNRAS.470.3283S,2017ApJS..228....1Z,2018A&A...619A.159M,2019A&A...627A.159H,2020A&A...634A.112T}).

The external emission field  is indicated via an isotropic blackbody with the temperature $T=h\nu_{\rm p}/(3.93k_B)$,
where the maximum frequency of seed photons in the ${\nu}-{\nu}{F}_{\nu}$ space is denoted by $\nu_{\rm p}$ (see \citealt{2016MNRAS.461.1862K,2017ApJ...837...38K} for the details),
$h$ is Planck Constant, $k_B$ is Boltzmann Constant.  
Consider that the $\gamma$-ray radiation area is placed out of the BLR and in the dusty torus. 
Thus, the energy density value of the BLR $U_{\rm BLR}$ is reduced quickly, 
while the energy density value of the dusty torus $U_{\rm torus}$ is approximately kept unchanged (see \citealt{2008MNRAS.387.1669G,2009MNRAS.397..985G}).
Accordingly, in this model, the $U_{\rm BLR}$ is considered as an arbitrary free parameter 
while the $U_{\rm torus}=3\times10^{-4} \Gamma^{2}$ erg cm$^{-3}$ (e.g., \citealt{2007ApJ...660..117C}) is chosen in the jet co-moving frame.
Furthermore, the influences of Klein-Nishina and the self-absorption (e.g., \citealt{1979rpa..book.....R,1970RvMP...42..237B}) in the inverse Compton scattering and synchrotron radiation are
appropriately assumed in this work, respectively  (see \citealt{2016MNRAS.461.1862K,2017ApJ...837...38K} for the details). 
In the SED modeling of {Figure~\ref{figSED1}}, the model forecasting is assumed as 
the inherent radiation.
It is then compared with the observational information 
(e.g., \citealt{2011ApJ...728..105Z,2013ApJ...764..113Z,2013MNRAS.431.2356Z,2014MNRAS.442.3166Z,2016A&A...585A...8Z,2012ChA&A..36..115K,2014ApJS..215....5K,2014JApA...35..385K,2016MNRAS.461.1862K,2017ApJ...837...38K}).

To constrain the model parameters and obtain a combination of their optimal values, a {\it McFit}  procedure 
is employed to fit the quasi-simultaneous multi-waveband SED of 3C 454.3 that was given in \cite{2009ApJ...699..817A}. 
The {\it McFit} procedure is a fitting tool that adopts a Bayesian Monte Carlo (MC) fitting technique for the confident 
fitting of parameters limited by the data even in the presence of other unconstrained parameters (see \citealt{2015ApJ...806...15Z,2016ApJ...816...72Z}). 
The mentioned technique could derive the optimal fitted parameters 
and their corresponding uncertainties through the converged MC chains (see \citealt{2015ApJ...806...15Z,2016ApJ...816...72Z} for the details).

\section{Modeling the SED of 3C 454.3}\label{section3}

\begin{figure*}
\centering
\includegraphics[width=8.5cm,height=6.5cm]{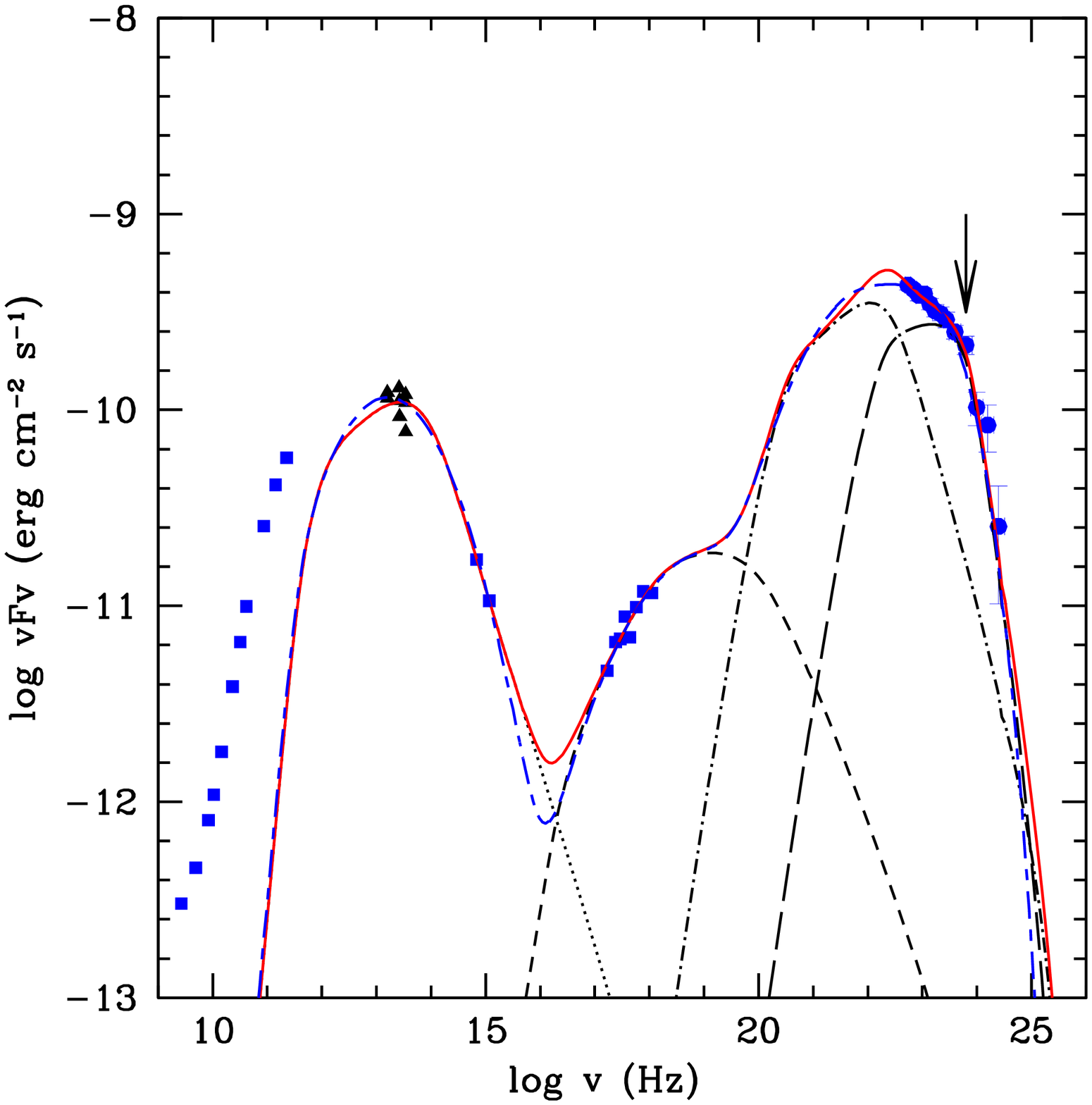}
\includegraphics[width=8.5cm,height=6.5cm]{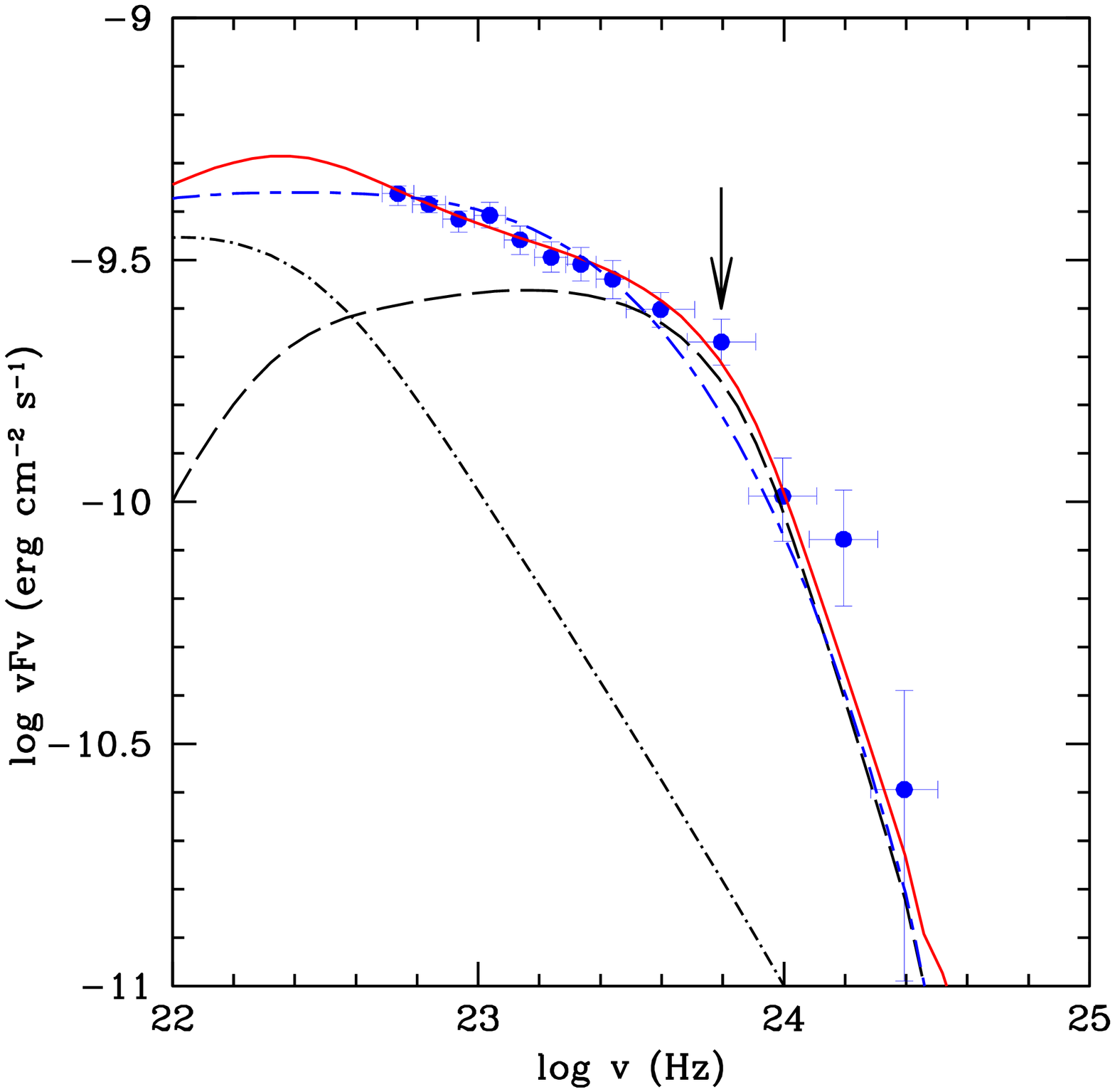}
\caption{The SED of 3C 454.3 (obtained from \citealt{2009ApJ...699..817A}
and reproduced by permission of the AAS).
The synchrotron, SSC, $\rm EC_{torus}$, $\rm EC_{BLR}$,  
and total emission based on the broken power-law electron spectrum
are indicated by dotted, dashed, dot-dashed, long-dashed, 
and solid lines, respectively. 
The total emission using the  log-parabolic electron spectrum is indicated by long-dashed-dashed (blue) line.
The right panel is zoomed in {\it Fermi}-LAT spectrum. }
\label{figSED1}
\end{figure*}

The simultaneous multi-wavelength data acquired from high energy $\gamma$-rays ({\it Fermi}-LAT), X-ray, UV and optical (Swift satellite)
and radio (IRAM 30 m and Effelsberg 100 m telescopes) for 3C 454.3 extracted within the time interval MJD 54685-54690 (2008	 August 7-12) 
and non-simultaneous mid-IR (infrared radiation) data acquired with the VLT/VISIR instruments in the early 2008 July
are gathered from \cite{2009ApJ...699..817A} and presented in Figure~\ref{figSED1}. 
In the Fermi band of Figure \ref{figSED1}, a spectral break is apparent near 2 GeV, with photon indices
of $\Gamma_1 \simeq 2.3 \pm 0.1$ and $\Gamma_2 \simeq 3.5 \pm 0.25$ lower and higher than the break, respectively \citep{2009ApJ...699..817A}.

The one-zone jet model is employed as described in Section \ref{section2}
to regenerate the multi-waveband SED of 3C 454.3. 
In this model, there exist 10 free parameters in the Syn + SSC + EC (BLR) + EC (torus) model:
$B$, $\delta$, $R$, $p_{1}$, $p_{2}$, $\gamma_{\rm min}$, $\gamma_{\rm max}$, $\gamma_{\rm b}$, $N_{0}$ {and} $U_{\rm BLR}$ for the broken power-law distribution electron spectrum.
To decrease the number of these free parameters, the emitting area radius in the jet
frame could be bounded with the minimum variability timescale and redshift with $R \leqslant\delta c t_{\rm var}/(1+z)\sim 1.40 \times 10^{15} \delta$ cm,
where the intra-day variation of 
 Fermi $\gamma$-ray band was determined in 3C 454.3 
(e.g., \citealt{2019ApJ...875...15W,2020ApJS..248....8D}).
In addition, a typical $\gamma$-ray variability timescale in days was reported in other literature (e.g., \citealt{2018RAA....18...40Z}). 
A 1-day conservative estimation is assumed and employed in this work. 
The maximum electron Lorentz factor is assumed in this work to be $\gamma_{\rm max}=1\times10^{8}$ 
($\gamma_{\rm max}{\gg}100\gamma_{\rm b}$) 
and does not have any considerable effect on the fundamental results (e.g., \citealt{2016MNRAS.461.1862K}). 
The next eight parameters, including $B$, $\delta$, $\gamma_{\rm min}$, $p_{1}$, $p_{2}$, $\gamma_{\rm b}$, $N_{0}$
and $U_{\rm BLR}$, 
are left free in the fitting.

\begin{table*} 
\centering
\caption{The relevant parameters of 3C 454.3 (Best-fit model and output parameters).}\label{Tab1}  
\begin{tabular}{lccclcc}
\hline \hline
\multicolumn{3}{c}{broken power-law electron spectrum}    &&\multicolumn{3}{c}{log-parabolic electron spectrum}     \\
\cline{1-3}  \cline{5-7} 
Parameter &  Parameters range  &  Best-fit model parameters         && Parameter &  Parameters range  &  Best-fit model parameters      \\
\hline
$B${(G)}	                           	                           &[0.01, 3.85]            &$1.27_{-0.11}^{+0.03}$                     && $B${(G)}	 &[0.01, 3.85]            &$1.21_{-0.07}^{+0.04}$       \\
$\delta$  	                           	                                                 &[10.0, 47.0]            &$22.92_{-0.36}^{+0.53}$                 && $\delta$  	                       &[10.0, 47.0]            &$23.62_{-0.33}^{+0.44}$     \\    
$p_{1}$  	                           	                                                 &[1.20, 3.50]            &$2.63_{-0.10}^{+0.03}$                   && $s$  	                        &[0.2, 5.90]            &$2.13_{-0.29}^{+0.45}$       \\
$p_{2}$  	                           	                                                 &[3.10, 7.86]           &$4.92_{-0.07}^{+0.10}$                    && $b$  	                           	  &[0.01,2.96]            &$1.53_{-0.04}^{+0.02}$       \\
$\gamma_{\rm min}$    	     	                                                 & [1.0, 200.0]           &$118.57_{-8.20}^{+5.35}$                 && $\gamma_{\rm min}$            	  &[1.0, 200.0]            &$54.31_{-38.28}^{+14.78}$   \\
$\gamma_{\rm b} (10^{2})$  	                           & [1.0, 100.0]             &$9.89_{-0.39}^{+0.67}$                   && $\gamma_{\rm 0} {(10^{2})}$        & [1.0, 100.0]        &$2.85_{-0.66}^{+0.83}$   \\
$N_{0} (10^2~{\rm cm}^{-3})$	                        & [0.01, 15000.0]       &$2036.38_{-1770.43}^{+410.38}$          && $K_{0}(10^{-5}~{\rm cm}^{-3})$    & [0.01, 15000]     &$70.89_{-26.91}^{+67.20}$   \\
$U_{\rm BLR} (10^{-5}{\rm erg~cm}^{-3})$ 	     & [0.01, 200.0]           &$46.05_{-13.52}^{+5.30}$                 &&  $U_{\rm BLR} (10^{-5}{\rm erg~cm}^{-3})$ 	     & [0.1, 200.0]   &$35.30_{-4.57}^{+7.73}$ \\
\hline
$\chi^2$/dof 	                           	                                         &                             &76.3/23                                     && &&     84.6/23  \\
$r_{\rm diss}$  	                                                              &                             & $\sim4.00R_{\rm BLR}$ ($\sim0.78 {\rm pc}$)   \\
\hline \hline
\end{tabular}\\
\end{table*}

\begin{figure*}
\includegraphics[width=17.5cm,height=14.3cm]{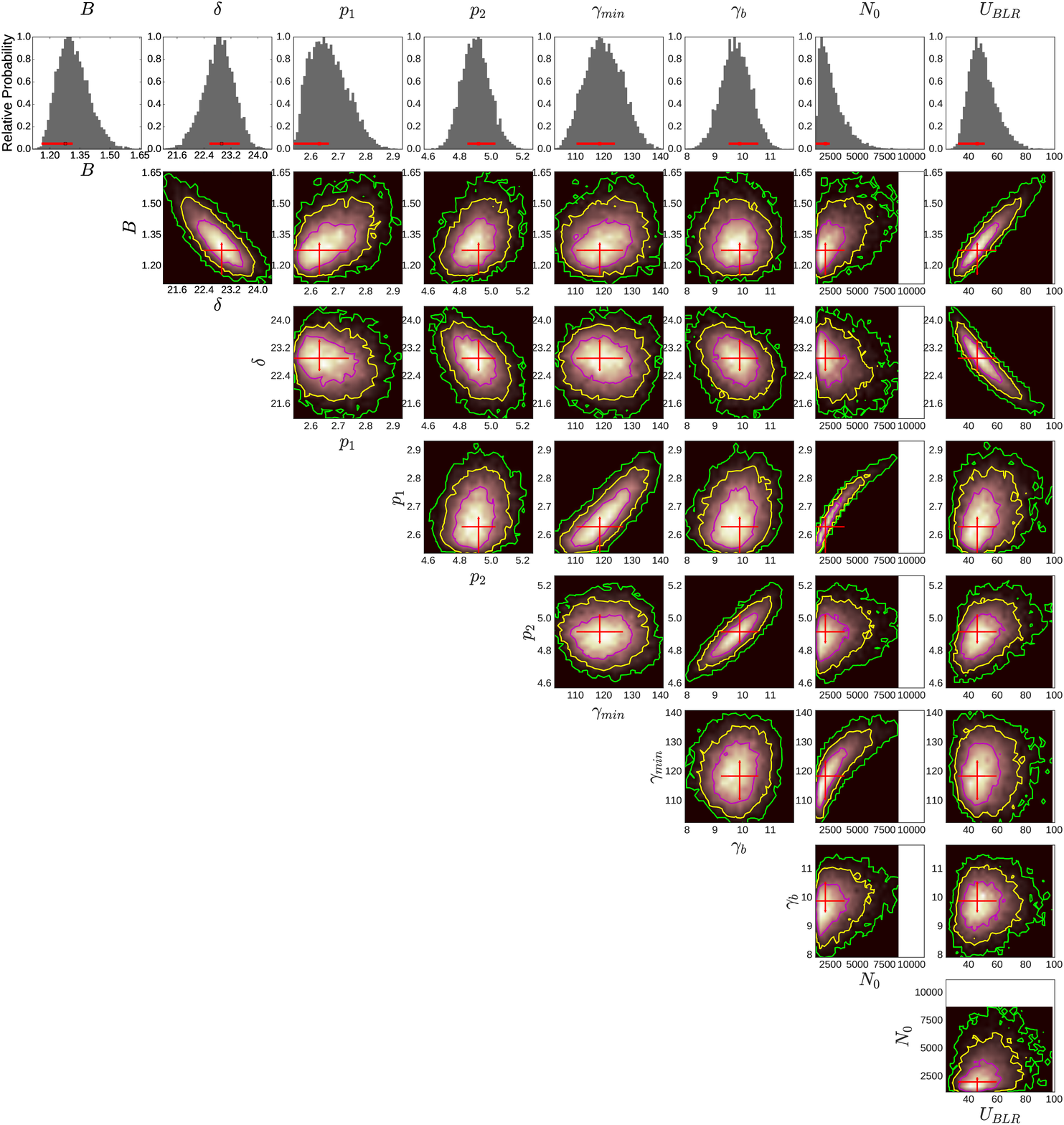}
\caption{Parameter constraints of our model fitting. Histograms and contours show the likelihood map of the parameter-constraint outputs by the {\it McFit} code. 
Red crosses mark the best-fit values and 1$\sigma$ error bars. All parameters are constrained in reasonable ranges (see Table \ref{Tab1}).}
\label{figparerr}
\end{figure*}

In SED fitting, 23 observational data points could be found, including 2 UV, 8 X-ray, and 13 $\gamma$-ray data points.  
The observational uncertainties corresponding to the data samples in the $\gamma$-ray bands are chosen in the {\it McFit} fitting.
Unfortunately, since uncertainties are not  available for the UV and X-ray data points acquired by  \cite{2009ApJ...699..817A}, 
two percent of the observational flux is considered as the systematic flux error (e.g., \citealt{2014A&A...567A.135A,2016MNRAS.461.1862K}).
Now, SED fitting is accomplished according to the {\it McFit} fitting technique within the permissible range of parameters (see Table \ref{Tab1}). 
The obtained results of the optimal fitting are indicated in Figure \ref{figSED1}. 
A optimal combination of model parameters is extracted by the {\it McFit} fitting technique.
The obtained best-fit model parameters, as well as uncertainties, are presented in Table \ref{Tab1}. 
The distributions of parameter values for 3C 454.3 are shown in Figure \ref{figparerr},
where the likelihood map of the parameter-constraint outputs by the {\it McFit} code are indicated by histograms and contours.
The colored contour lines represent a two-dimensional contour map of the relative marginalization probabilities.
The red crosses indicate the optimal fitted values and 1$\sigma$ error bars.

It should be noted that the GeV spectral break of 3C 454.3 could be entirely reproduced through the leptonic jet model with 
the Syn + SSC + EC (BLR) + EC (torus) model. 
The value distribution of all obtained model parameters is usually displayed as a (or approximately) normal distribution (see Figure \ref{figparerr}).
However, the optimal fitted values and uncertainties obtained from the optimal combination of model parameters for all model parameters are not always related to peak probabilities in a one-dimensional histogram or two-dimensional contour plot of relative marginalization probabilities 
(see \citealt{2015ApJ...806...15Z}). For instance, the best fit of the density normalization coefficient $N_0$ and the indices $p_1$ of electron spectrum in particular appear to be on the edge of the considered region.
As could be seen from Table \ref{Tab1}, all the modeling parameters lie within acceptable ranges. 
For instance, almost all fitted model parameters are consistent with those obtained with other previous works. 
For example, 
the Doppler factor $\delta\simeq {22.92_{-0.53}^{+0.36}}$ is compatible with the Doppler coefficient of
$\delta=\left[\Gamma\left(1-\beta\cos\theta\right)\right]^{-1}\approx 22.6$ \citep{2019ApJ...875...15W} 
obtained from a study on 43 GHz Very Long Baseline Interferometry
(VLBI) images using the method proposed by \cite{2005AJ....130.1418J,2017ApJ...846...98J}
with a bulk Lorentz coefficient $\Gamma = 20.4 \pm 0.4$
and viewing angle of the K16 knot path corresponding to the sightline
$\theta=2^o.5\pm 0^o .3$ \citep{2019ApJ...875...15W}.
The minimum electron Lorentz coefficient $\gamma_{\rm min} = {118.57_{-8.20}^{+5.35}}$, 
which corresponds to a specific situation of the electrons pre-shocked (e.g., \citealt{1998ApJ...497L..17S}), 
is compatible with values from the literature
(e.g., \citealt{2014ApJ...788..104Z,2014ApJS..215....5K,2019NewAR..8701541H}). 
However, the BLR energy density of 3C 454.3 ($U_{\rm BLR} = ({46.05_{-13.52}^{+5.30}}) \times 10^{-5}~{\rm erg~cm^{-3}}$ in the rest frame) in our modeling is about two orders of magnitude 
less than that of luminous FSRQs, where $U_{\rm BLR} \sim 2.6\times 10^{-2}$ ${\rm erg~cm^{-3}}$ due to the emission location
(see \citealt{2008MNRAS.387.1669G,2009MNRAS.397..985G} for details).

\section{Conclusion and Discussion}\label{section4}

In the context, the one-zone homogeneous leptonic jet model [Syn+SSC + EC (BLR) + EC (torus)] and {\it McFit} technique are employed 
to fit the quasi-simultaneous multi-waveband SED of 3C 454.3, 
where two EC components are considered in the fitting, namely, both Compton-scattered BLR and dust torus radiations. 
Combining two EC components provides an appropriate fitting to the quasi-simultaneous multi-waveband SED of 3C 454.3
and explains the GeV spectral break of 3C 454.3. 
The GeV spectral break of 3C 454.3 can be reproduced naturally using the two EC components ($\rm EC_{BLR}$ + $\rm EC_{torus}$) jet model, 
considering a broken power-law distribution electron spectrum and Klein-Nishina effect. 
We argue that the GeV spectral break of 3C 454.3 may originate from an inherent break in the energy distribution of the emitted particles and the Klein-Nishina effect.
However, although both the electron distribution break and the Klein-Nishina affect the resulting shape of the $\gamma$-ray spectrum (e.g., \citealt{2018ApJ...853....6L,2019ApJ...884..116L}), detecting the dominant factor based on the current work is complicated and should be further verified in the next work.
Based on the energy densities of the BLR external photon fields and the scattering distance from the central black hole, the position of
the $\gamma$-ray emitting area of 3C 454.3 could be firmly constrained between the BLR and the dust torus (outside BLR, inside dust torus)
(the scattering distance of the $\gamma$-ray radiation area from the central black hole $r_{\rm diss}$ is approximately several times of $R_{\rm BLR}$).

\subsection{$\gamma-$ray emission region of 3C 454.3}\label{section41}

The $U_{\rm BLR}$ could  be utilized to estimate the position of $\gamma$-ray emission area (see the work performed by \citealt{2014ApJ...782...82D}). 
Moreover, the $U_{\rm BLR}$ can be estimated
from the luminosity of BLR and the BLR radius 
estimated from the accretion disk luminosity 
 (e.g., \citealt{2008MNRAS.387.1669G}).
For 3C 454.3, \cite{2011MNRAS.410..368B} reported the luminosity of
the accretion disk $L_{\rm disk}  \simeq {6.75} \times 10^{46}~{\rm erg~s^{-1}}$ 
and the size of the BLR $R_{\rm BLR} \simeq  6 \times 10^{17}$ cm ($\simeq 8 \times 10^3 R_G$)
obtained from the relation of \cite{2007ApJ...659..997K}. 
Based on the relationship between the $U_{\rm BLR}$ and $R_{\rm BLR}$ \citep{2008MNRAS.387.1669G}, $U_{\rm BLR} \simeq 5 \times 10^{-2}~{\rm erg~cm^{-3}}$  can be  estimated, which is consistent with that of luminous FSRQs (e.g., \citealt{2008MNRAS.387.1669G,2014ApJ...782...82D}).
The fitting value of $U_{\rm BLR}$ (see Table \ref{Tab1}) derived from the modeling procedure
is two orders of magnitude less than the mentioned estimated amount. 
This implies that a radiating area is located outside the BLR's outer boundary, 
which is compatible with the predicted positions for the $\gamma$-ray radiation region in other sources 
(e.g., \citealt{2014ApJ...782...82D,2016ApJ...821..102B,2017ApJ...837...38K,2017ApJS..228....1Z}).

The relation between the $U_{\rm BLR}$ and the dissipation distance $r_{\rm diss}$ from the central black hole
(e.g., \citealt{1996MNRAS.280...67G,2009MNRAS.397..985G})
could be approximated as \citep{2009ApJ...704...38S,2012ApJ...754..114H}
\begin{equation}\label{eq3}
  U_{\rm BLR}(r)=\frac{\tau_{\rm BLR} L_{\rm disk}}{4\pi R^2_{\rm BLR}c[1+(r_{\rm diss}/R_{\rm BLR})^3]},\
\end{equation}
where $\tau_{\rm BLR}$  is a portion of the disc luminosity reprocessed into BLR emission, 
where its typical value is $\tau_{\rm BLR}=0.1$ (e.g., \citealt{2014Natur.515..376G}).
The calculated distance from the central black hole to the radiating  blob of 3C 454.3 denoted 
by $r_{\rm diss}$ can be calculated as 
{$r_{\rm diss} \simeq 4.00 R_{\rm BLR} \simeq 0.78$ pc ($\simeq 3.20 \times 10^4 R_G$}), 
using the $U_{\rm BLR}$ derived from the model fitting the SEDs  of 3C 454.3. 
This is compatible with previously published work in several blazars 
(e.g., \citealt{2010ApJ...714L.303F,2014ApJ...782...82D,2015MNRAS.454.1310Y,2015ApJ...803...15P,2017ApJ...837...38K,2017ApJS..228....1Z}).

\subsection{$\gamma-$ray origin of 3C 454.3}\label{section42}
In this work, 
the SED (GeV spectral break) of 3C 454.3 could be well reproduced via the leptonic jet model with the Syn+ SSC+ EC (BLR) + EC (torus) model using the electron spectrum with a broken power-law distribution. 
\cite{2013ApJ...771L...4C} suggested that the GeV spectral break could be produced by the EC dissipation of photons from the dusty torus and BLR, 
assuming an electron spectrum represented with a log-parabolic function.
The different electron spectra are related to different physical origins. 
A log-parabolic electron spectrum is produced by stochastic acceleration process (Fermi II process) in the jet 
(see \citealt{2004A&A...413..489M,2004A&A...422..103M,2006A&A...448..861M}), 
where the acceleration dominates over radiative cooling (e.g., \citealt{2009A&A...501..879T,2011ApJ...739...66T}).
{\bf\color{red} }
However, the electron spectrum with a broken power-law distribution is a steady-state electron spectrum, where no acceleration process in the jet's emitting region.
There is a balance between the emitting particle cooling and escape rates in the emitting region
(e.g., \citealt{1962SvA.....6..317K}; \citealt{1994ApJ...421..153S}; \citealt{1996ApJ...463..555I}; \citealt{1998A&A...333..452K}; \citealt{1998MNRAS.301..451G}; \citealt{2002ApJ...581..127B}; \citealt{2012ApJ...748..119C}; \citealt{2013ApJ...768...54B}).
The broken energy $\gamma_{\rm b}$ of the broken power-law electron spectrum is a critical energy due to the balance between the emitting particle escape and cooling during the radiation process, where less than this critical energy, the energy loss process is dominated by particle escape, and above the critical energy, the energy loss is dominated by cooling (e.g., \citealt{2013ApJ...763..134F,2018MNRAS.478.3855Z} and reference in).
Moreover, the shock acceleration shapes a power-law particle distribution (\citealt{1949PhRv...75.1169F}). 
Mixing of shocked populations may cause the break, or the broken-power law could approximate a more precise functional form.

In order to distinguish these two electron spectra, in this work, the fitting results are compared using the two electron spectra with a broken power-law distribution and a log-parabola distribution. 
The obtained best-fit model parameters, as well as uncertainties, are also presented in Table \ref{Tab1}.
In addition to the electron spectrum parameters, other jet model parameters (e.g., $B$, $\delta$, $U_{\rm BLR}$)
using a log-parabolic electron spectrum are consistent with that of the electron spectrum with a broken power-law distribution (see Table \ref{Tab1}).
However, the fitted chi-square {($\chi^2$/dof =84.6/23)} of the log-parabolic electron  spectrum is larger than that {($\chi^2$/dof =76.3/23)} of the electron spectrum with a broken power-law distribution.
These scenarios suggest that the broken power-law distribution electron spectrum fits the SED better. 
In addition, in Figure \ref{figSED1}, the SED fitting result (total emission) using the log-parabolic electron spectrum is indicated by the long-dashed-dashed (blue) line,
which exhibits a strong curvature in the Fermi GeV $\gamma$-ray band. 
In terms of intuitive phenomenology, the curvature-type radiation particle distribution may be more inclined to produce a curvature-type radiation photon spectrum (see \citealt{2004A&A...413..489M,2004A&A...422..103M,2006A&A...448..861M,2020ApJ...898...48A}); 
the broken-type radiation particle distribution may be more inclined to produce a broken-type radiation photon spectrum. 
Here, the curvature spectrum cannot effectively produce the $\gamma$-ray spectrum break (see Figure \ref{figSED1}).
On the contrary, a broken power-law distribution electron spectrum can effectively generate the $\gamma$-ray spectrum break.
Based on the fact that the observed spectrum is broken, not a curvature one, 
a broken power-law electron spectrum, which may be more effectively present the observation characteristics. 
Further study is required to verify whether this interpretation is suitable for all GeV spectral breaks.

We should note that the GeV spectral break may be caused by an inherent break in the emitted electron spectrum,
where the break energy can be determined by an inherent break in the emitted electron spectrum.  
By modeling SED 3C 454.3, the break energy {$\gamma_b=9.89_{-0.39}^{+0.67}\times 10^2$} of the electron spectrum and {$\Gamma (\simeq \delta) =22.92_{-0.36}^{+0.53}$} are obtained. 
Assuming the most prominent line of the BLR cloud is derived from the Ly-$\alpha$ line, 
the spectrum is considered as a blackbody with an approximated maximum $\nu_{\rm ext} \simeq 2\times10^{15}~\Gamma$ Hz (e.g., \citealt{2008MNRAS.387.1669G}).
However, it should note that 
the most prominent line from the perspective of the gamma-ray emitting region may vary by position with respect to the ionization gradient in the BLR due to the relativistic effects (see \citealt{ 2016ApJ...830...94F}), which may affect the blackbody spectrum.
According to the Thomson regime or KN regime, the EC maximum frequency 
\begin{equation}
\nu_{\rm ec, thom}^{\rm p} \simeq \frac{4}{3}\gamma_{\rm b}^2\Gamma\nu_{\rm ext}{\frac{\delta}{1+z}}  \simeq 69.98~{\rm GeV} 
\end{equation}
or 
\begin{equation}
\nu_{\rm ec, kn}^{\rm p} \simeq \frac{2}{\sqrt{3}} \frac{\gamma_{\rm b}m_e {c^2}}{h} {\frac{\delta}{1+z}}  \simeq 7.20~{\rm GeV} 
\end{equation}
can be obtained, 
where $m_e$ is the electron mass (e.g., \citealt{ 2018ApJS..235...39C} and references therein). 
The EC peak frequency $\nu_{\rm ec, kn}^{\rm p} \simeq $ {7.20} GeV in the KN regime is consistent with the break energy of 3C 454.3 (a spectral break around 2.4 GeV).  
This may imply that the spectral break of 3C 454.3 may originate from an inherent break in the energy distribution of the emitted particles in the case of the KN regime.

We also should note that the Klein-Nishina influences on Compton scattering of the BLR target photons, 
where the dominant BLR emission line (assuming about 10.2 eV Ly-$\alpha$ line, e.g., \citealt{2010ApJ...721.1383A}; \citealt{2013ApJ...771L...4C}) with the energy  $E_{\rm seed,BLR}$, 
generally creates a break ($E_{\gamma}$) at a few GeV (e.g., \citealt{2009ApJ...699..817A,2010ApJ...712..238F}),
\begin{equation}  
E_{\gamma} \simeq E_{\rm seed,BLR}\gamma_{\rm b}^2\Gamma\frac{\delta}{1+z}.
\end{equation}
However, whether the effect of the electron distribution break or the Klein-Nishina effect on the resulting shape of the $\gamma$-ray spectrum dominating on the physical process is beyond the scope of this work. We leave these issues in the future.

We should note that, in this work, we only use the two external photon fields jet model to fit the steady-state SED of 3C 454.3 and do not consider fitting SED (e.g., flare) of other states of 3C 454.3 or other GeV-break sources. Based on 3C 454.3 steady-state energy spectrum simulation, we propose and test a possible origin of GeV spectral break that the spectral break (GeV break) of 3C 454.3 may originate from both an inherent break in the energy distribution of the emitting particles and Klein-Nishina influence (see \citealt{2009ApJ...699..817A}  for the related discussions).  As future work, this explanation's applicability to all GeV-break blazars and all states should be more studied and tested.

\section*{Acknowledgements}
We thank the anonymous editor and referee for very constructive and helpful comments and suggestions, which greatly helped us to improve our paper.
This work is  partially supported by the National Natural Science Foundation of China  (Grant Nos.11763005, 11873043, U1931203, U1831138), 
the Science and Technology Foundation of Guizhou Province (QKHJC[2019]1290)
and the Research Foundation for Scientific Elitists of the Department of Education of Guizhou Province (QJHKYZ[2018]068).
\section*{Data availability}
The data underlying this article will be shared on reasonable request to the corresponding author.
\bibliographystyle{mnras}
\bibliography{mnras} 
\bsp	
\label{lastpage}
\end{document}